\documentclass[a4paper,11pt]{article}
\usepackage{pos}
\usepackage{amsmath}
\usepackage{caption,subcaption}
\usepackage{braket}
\usepackage{slashed}
\usepackage{bbold}
\usepackage{graphicx}
\usepackage {graphpap}
\usepackage{tikz}
\usepackage{fontenc}
\usepackage{times}
\usepackage{mathptmx}
\usepackage[utf8]{inputenc}

\title{Running HMC Simulation with Python via QUDA}

\author*[a]{Shuhei Yamamoto}
\author[a]{Simone Bacchio}
\author[a]{Jacob Finkenrath}

\affiliation[a]{Computation-based Science and Technology Research Center, The Cyprus Institute, Cyprus}

\emailAdd{s.yamamoto@cyi.ac.cy}

\abstract{Lyncs-API is a Python API for Lattice QCD applications.  It is designed as a Python toolkit that allows the user to use and run various lattice QCD libraries while programming in Python.  The goal is to provide the user an easy programming experience without scarifying performance across multiple platforms, by preparing a common framework for various softwares for lattice QCD calculations.  As such, it contains interfaces to, e.g., c-lime, DDalphaAMG, tmLQCD, and QUDA.  In this proceeding, we focus on a Lyncs interface to QUDA, named Lyncs-QUDA, and present a small tutorial on how to use this Python interface to perform a HMC simulation using QUDA.}

\FullConference{%
The 39th International Symposium on Lattice Field Theory,\\
8th-13th August, 2022,\\
Rheinische Friedrich-Wilhelms-Universität Bonn, Bonn, Germany
}

\begin{document}
\maketitle

\section{Introduction}
\begin{figure}[hbt]
\centering
\includegraphics[scale=0.3]{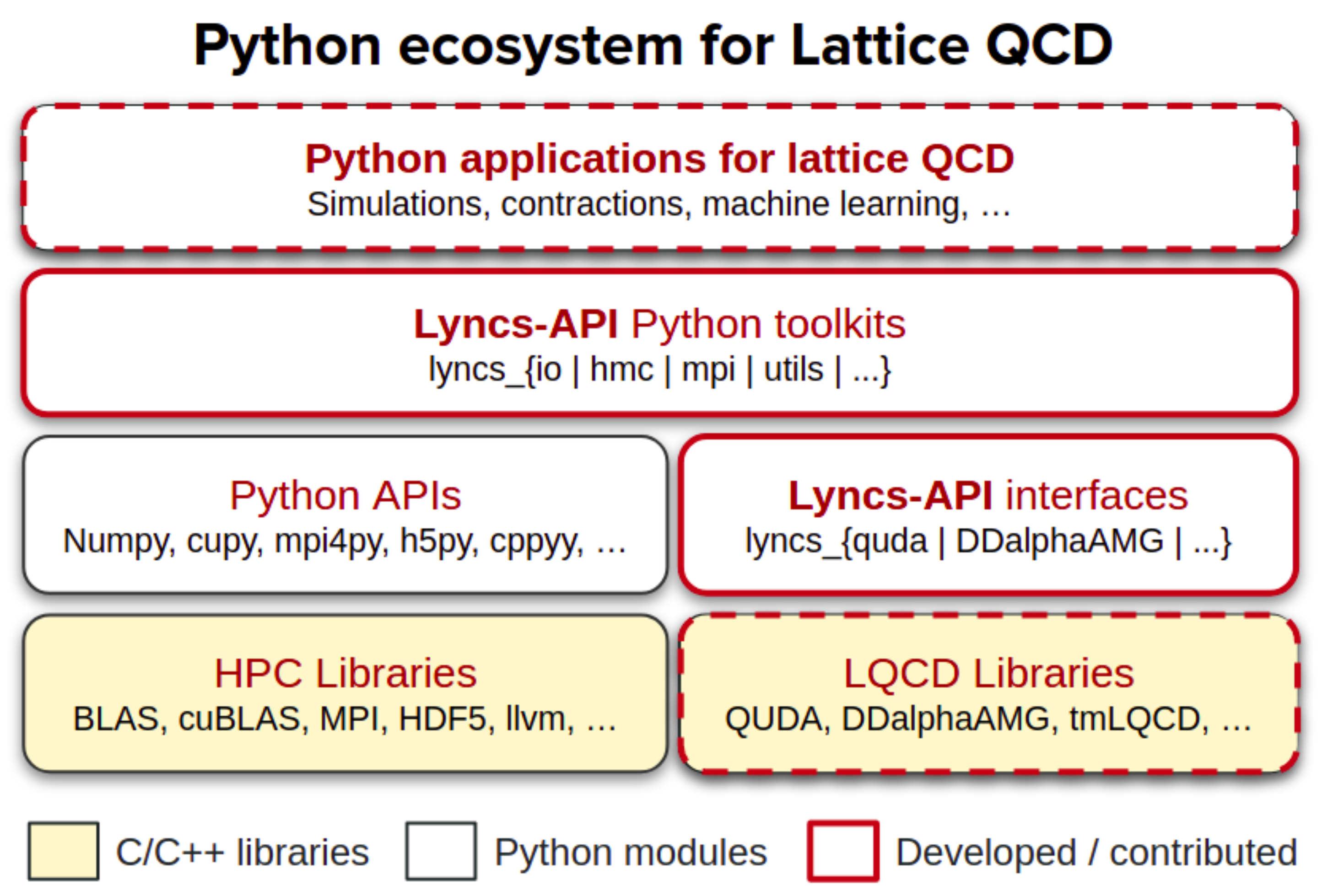}
\caption{A schematic diagram showing how Lyncs-API is organized.}
\label{Fig:Lyncs-API}
\end{figure}

Portability and productivity are new demands in High-Performance Computing (HPC) in addition to Performance, and lattice QCD applications are no exception.  While performance has been sought for in those applications, portability and productivity have also become important factors.  Lyncs-API is a response to those newly emerging trends and tries to achieve those three factors by developing a Python toolkit that allows the user to use and run various lattice QCD libraries while programming in Python \cite{Bacchio:2022bjk}.   For this purpose, it uses a number of Python packages and aims at implementing Pyhton interfaces to various lattice QCD libraries.  The structure of this API is shown schematically in Fig.~\ref{Fig:Lyncs-API}.
In this work, we introduce a Python interface to one of the QCD libraries, namely QUDA, which is a library for performing calculations in lattice QCD on graphics processing units (GPU's) \cite{Clark:2009wm}.  The interface is available online at \url{ https://github.com/Lyncs-API/lyncs.quda}.  It can be installed via pip from this GitHub repository.  In the following, we outline how this interface is implemented, provide an example for how it can be used for pure-gauge HMC simulations, show what it produces, and analyze its performance.

\section{Implementation}

Lyncs-QUDA is designed to provide an intermediate layer of Python interface to QUDA.  What this means is that the interface exposes methods and objects of QUDA by directly accessing them internally. 
  This allows the user to program in Python while accessing to various QUDA objects and functionality.  To achieve this, Lyncs-QUDA prepares wrapper classes for QUDA functions and objects while categorizing QUDA objects based on functionality.  
  
These wrappers are contained in the wrapper scripts located in the directory \texttt{lyncs\_quda}.   Among them, a script, \texttt{lib.py}, contains a class on top of which all other classes are built.  The class is named \texttt{QudaLib}, and its instance, \texttt{lib}. The \texttt{lib} object loads QUDA objects from the QUDA library and includes definitions from its header files using \texttt{cppyy} \cite{7836841}.  Other classes access to QUDA objects through the \texttt{lib} object.  

\begin{figure}
\centering
\includegraphics[scale=0.6]{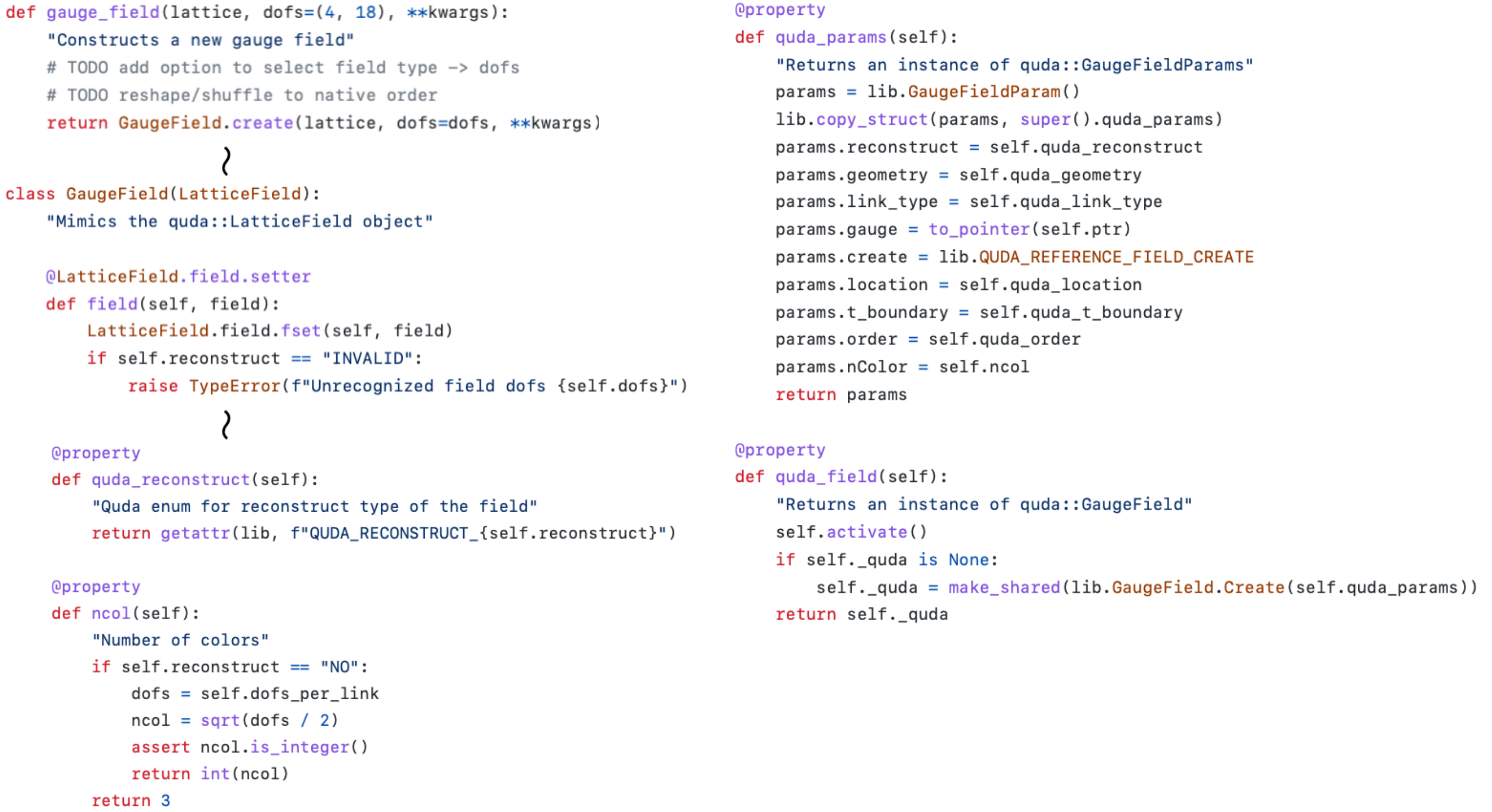}
\caption{A snapshot of a excerpt from \texttt{gauge\_field.py}, showing how wrapper scripts are implemented.} 
\label{Fig:GaugeImplementation}
\end{figure}

To give an idea how this is done, we take \texttt{gauge\_field.py} as an example to show how wrapper scripts are implemented.  An excerpt of the version of the script used for this proceeding is found in Fig.~\ref{Fig:GaugeImplementation}.  The scripts can start with some auxiliary functions, which in case of the \texttt{gauge\_field.py} are functions that return an instance of \texttt{GaugeField} class, which follows these auxiliary functions.  
This \texttt{GaugeField} class is the main object of this script.  It is modeled after the corresponding QUDA class, \texttt{quda::GaugeField}, with additional capabilities.  For instance, Python's \texttt{GaugeField} inherits from \texttt{LatticeField} class similarly to QUDA's \texttt{GaugeField} class and also offers some methods not found inside of \texttt{quda::GaugeField} such as those for computing various gauge actions and topological charges collected in this class.  In principles, there is a corresponding Python class for each QUDA class.
This Python \texttt{GaugeField} class stores the data field for a gauge field, represented either as a \texttt{numpy} or \texttt{CuPy} array.  It also manages meta-data such as the dimensions of the lattice, internal degrees of freedom, the number of colors, and reconstruction type.  To perform various operations through QUDA, it contains an instance of the \texttt{quda::GaugeField} class.  This is created and returned as a shared pointer based on the meta-data of the Python \texttt{GaugeField} class, supplied to the \texttt{Create} method of the \texttt{quda::GaugeField} class in the form of \texttt{quda::GaugeFieldParam}.  After the introduction of the corresponding QUDA object, the class definition offers various methods that rely on the QUDA object.  In the simplest case, those methods call the associated QUDA methods either through the internal QUDA object or via Python's \texttt{lib} object introduced earlier.  There are also other methods which combine several QUDA functions to perform more complicated tasks such as fermion force computation.  This is a basic design pattern of the wrapper scripts of this interface.  

On the GitHub page for this interface, there are also test and example scripts found in the respective directories.  Test scripts are to be used with \texttt{pytest} and allow the user to check if those wrapper scripts work as expected.  Example scripts show how these wrapper scripts can be used to solve practical problems.  We use one of the example scripts, named \texttt{hmc.py}, to illustrate how this interface can be used by applying it to pure-gauge HMC simulations, which is detailed in the next section.

\section{Example: pure-gauge HMC}
Lyncs-QUDA greatly simplifies programs for calculations of lattice QCD quantities requiring GPU's.  To illustrate it, we prepared a script for pure-gauge HMC simulations.  As a brief reminder, HMC simulation is performed in the following steps.
\begin{enumerate}
\item Pick an initial gauge configuration
\item Generate the conjugate momenta according to Gaussian distribution
\item Perform Molecular Dynamics (MD) integration
\item Accept/Reject a proposed configuration according to Metropolis criteria
\item Go back to (2)
\end{enumerate}
The steps from (2) to (5) are iterated as many times as needed to produce a required number of gauge configurations. 

To perform various tasks necessary to complete those steps, the HMC script is composed of several classes along with the usual main function.  These classes are \texttt{HMCHelper}, \texttt{Integrator}, and \texttt{HMC}.  Among those, the \texttt{HMCHelper} class does most of the work.  It generates the initial gauge configuration for Step (1) as well as a random conjugate momentum field with Gaussian distribution at each start of the MD integration, computes gauge action and force, and updates the gauge field and momentum.  The class, \texttt{Integrator}, uses this update method from the \texttt{HMCHelper} class to integrate over the MD trajectory.  This \texttt{Integrator} class provides the user several options for integration methods such as leap frog, MN2, or OMF4 \cite{Sexton:1992nu,2002PhRvE..65e6706O,Takaishi:2005tz}.  The \texttt{HMC} class combines these two classes to make a single HMC step, i.e., from Step (2) to (5), and lastly, the main function does initialization of Lyncs-QUDA and QUDA and then uses \texttt{HMC} to repeat HMC steps as many times.  
\begin{figure}[hbt]
\centering
\includegraphics[scale=0.6]{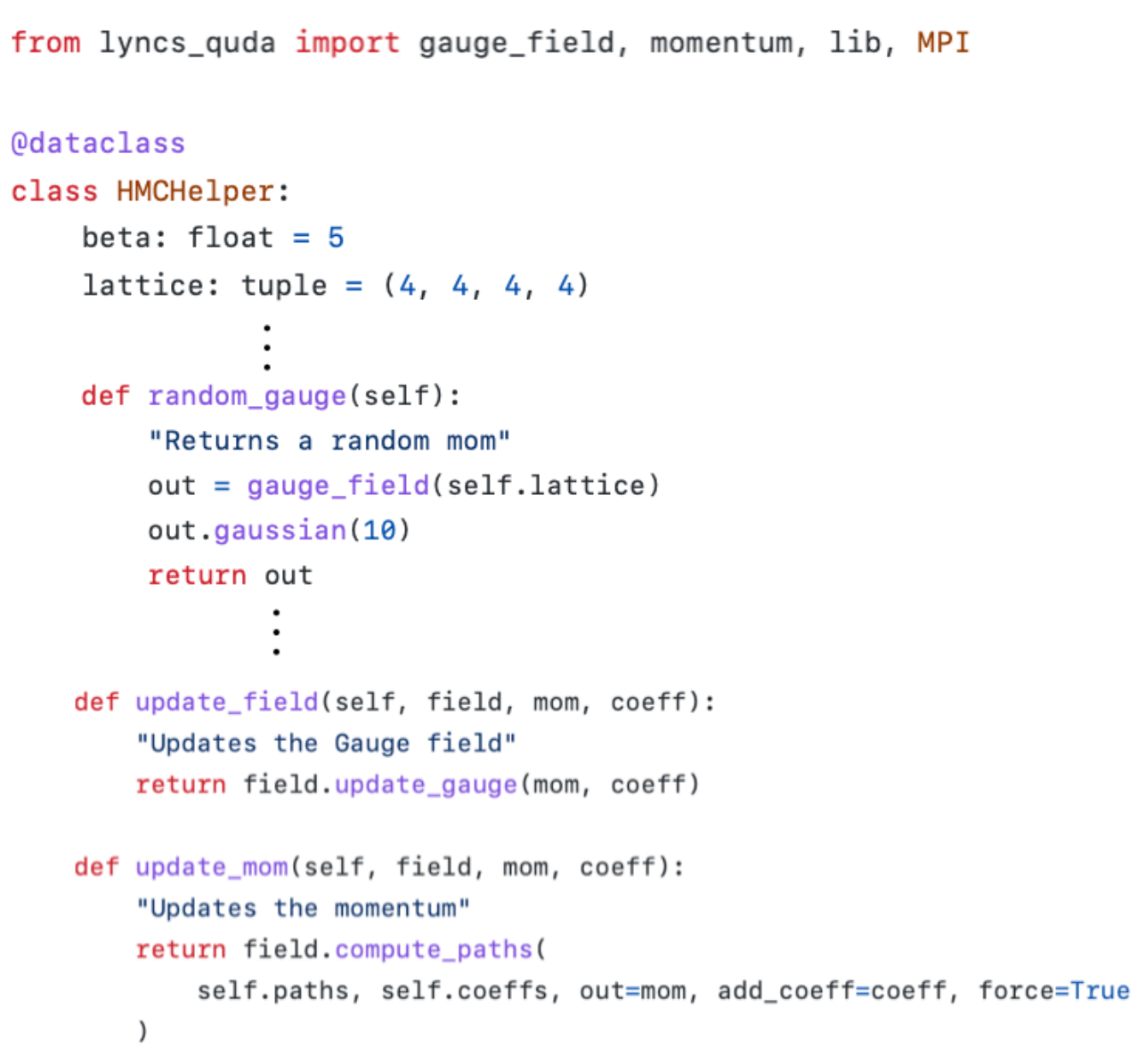}
\caption{A snapshot of an excerpt from \texttt{hmc.py} located in \texttt{examples} directory of the GitHub page.}
\label{Fig:HMCscript}
\end{figure}
As is mentioned, most of complicated tasks are performed by \texttt{HMCHelper}, which implements the core methods needed for HMC simulation.  Figure \ref{Fig:HMCscript} shows a snippet from the script.  As Fig.~\ref{Fig:HMCscript} shows, however, the code for these methods looks fairly simple.  
All of the implementation details are hidden inside of a few calls of related Lyncs-QUDA objects.
The user only needs to create relevant Lyncs-QUDA objects, and the required tasks are completed merely by calling methods of these Lyncs-QUDA objects.  In this manner, this Python interface significantly simplifies programming and lets the user focus on solving the problem instead of implementation of low level functions.  Furthermore, computationally intensive tasks are off-loaded to GPU's via QUDA while achieving programming simplicity at the same time.

\begin{figure}
\centering
\includegraphics[scale=0.4]{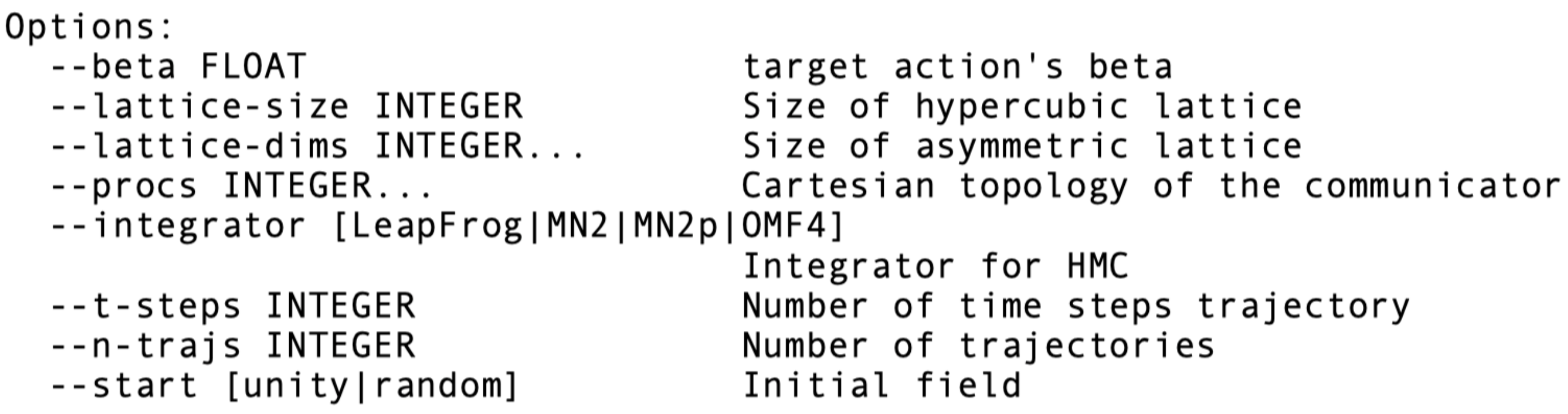}
\caption{Currently supported command-line options with help messages for the script.}
\label{Fig:HMCoptions}
\end{figure}

The example script for HMC simulations can be run from the command line with additional command-line options in the following way:
\begin{center}
\texttt{python examples/hmc.py --lattice-size 48  --beta 6.475.}
\end{center}
Currently supported options are shown in Fig.~\ref{Fig:HMCoptions}.  This script can be used to generate gauge configurations on various lattice sizes using variable number of processes specified as a Cartesian topology.  The user can change the value of $\beta$ and select different integration methods, the number of time steps per MD trajectory, and the number of MD trajectories.  We will add support for more options when the script grows into a programming suit of its own.

\section{Results}
In this section, we review the output of the HMC script and analyze the performance of the Lyncs-QUDA library.

\subsection{First Sanity Check}
\begin{figure}
\centering
\includegraphics[scale=0.6]{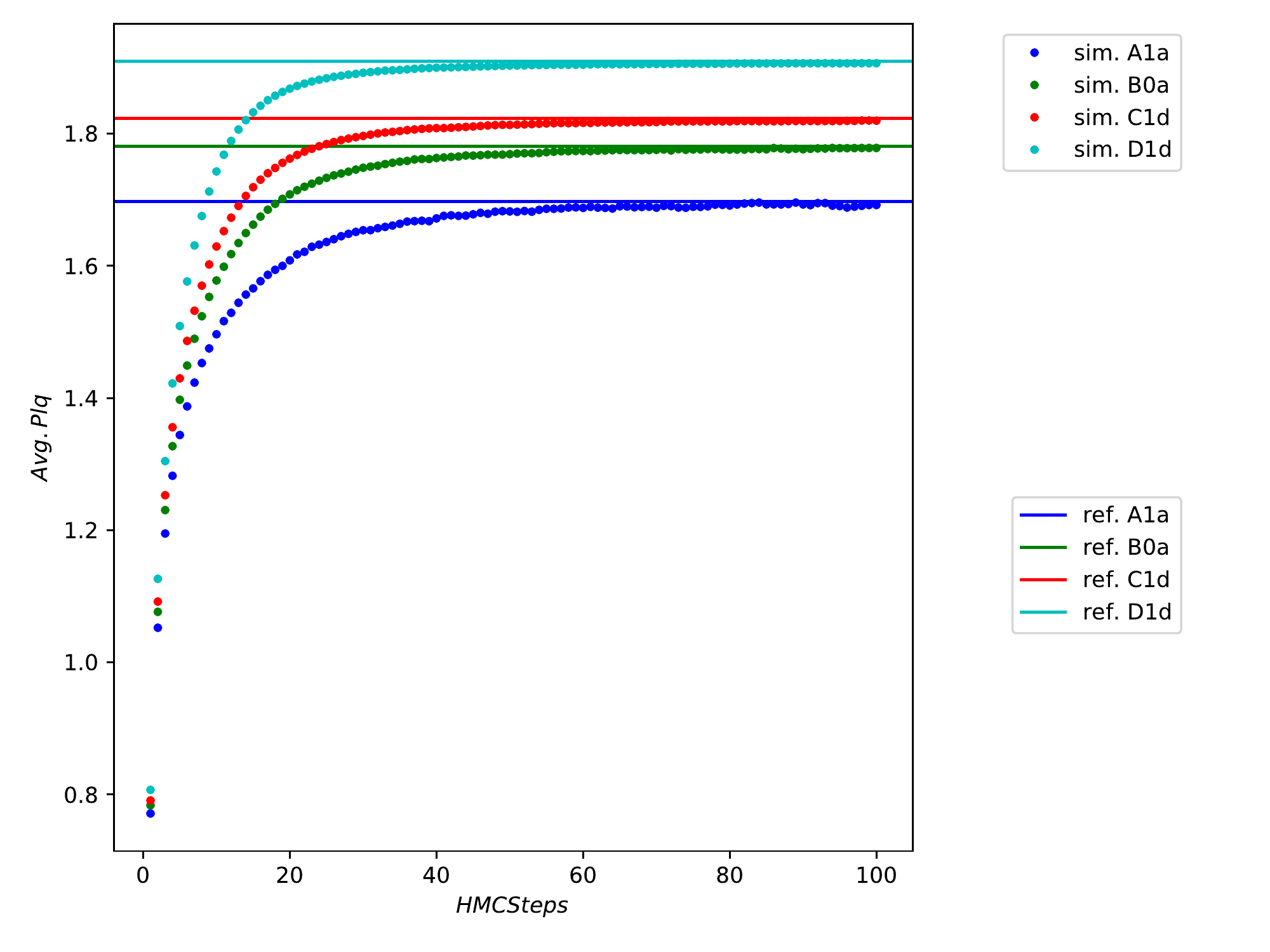}
\caption{A plot of average plaquette values on each configuration.  The dotted line corresponds to the values of the average plaquette on configurations generated by the HMC script, and the solid the thermalized average plaquette value taken from Ref.  The dotted line and solid lines of the same color correspond to the same ensemble.}
\label{Fig:AvgPlq}
\end{figure}

As the first sanity check, we have computed the value of average plaquette on each configuration generated by our HMC script and checked the thermalization process of the plaquette value.
These tests are performed on several lattices with different dimensions and $\beta$ values.  Thermalized values are taken from Ref.~\cite{Schaefer:2010hu}.  In Ref.~\cite{Schaefer:2010hu}, several observables are computed on various ensembles.  Among those ensembles, we have selected four ensembles with different lattice dimensions and $\beta$ values.  They are listed in Table~\ref{T:ensembles}.  
\begin{table}[bht]
\centering
\begin{tabular}{ |c|c|c|c| } 
 \hline
TAG & $\beta$ & L/a & T/a \\ 
 \hline
 $A1a$ & $5.789$ & 16 & 16 \\ 
 $B0a$ & 6 & 24 & 24 \\ 
  $C1d$ & 6.136 & 32 & 64 \\ 
   $D1d$ & 6.475 & 48 & 48\\ 
 \hline
\end{tabular}
\caption{Table of ensembles used for HMC simulation \cite{Schaefer:2010hu}.  Parameter values relevant to computation of plaquette values are listed.}
\label{T:ensembles}
\end{table}
Figure \ref{Fig:AvgPlq} shows how our HMC script thermalizes the plaquette values on each ensemble.  As is shown in the figure, our HMC script thermalizes the plaquette value to an appropriate thermealized value for each ensemble.  This means that our script works as expected.
We have also collected other metrics of the run such as system GPU and CPU memory usage per second, the value of the action, difference in the Hamiltonian on the current and proposed gauge configurations, and the exponent of the negative of the Hamiltonian at each HMC step.  Other metrics can be collected upon needs of the user with a minor modification of the script.  The collection and visualization of the metrics during the runs are done using an external Python package called "aim" \cite{Arakelyan_Aim_2020}.  This package allows the script to track runs, log information, and easily visualize and analyze the collected data dynamically through a web browser.  

\subsection{Profiling}
In this section, we analyze the performance of our script.  This analysis is done on a local machine, \textit{cyclamen}, which is equipped with Intel\textregistered Xeon\textregistered Gold 6130 and two Nvidia's Tesla P100, located at The Cyprus Institute.  The ensemble, D1d, with its lattice size of $48^4$ was chosen.  For profiling and visualization, we have used a Python module, \texttt{cProfile} to collect profiling data and an external package, \texttt{SnakeViz} for visualization and analysis \cite{snakeviz,10.5555/1593511}.  For this profiling, we have performed 100 HMC steps.

\begin{figure}
\centering
\includegraphics[scale=0.45]{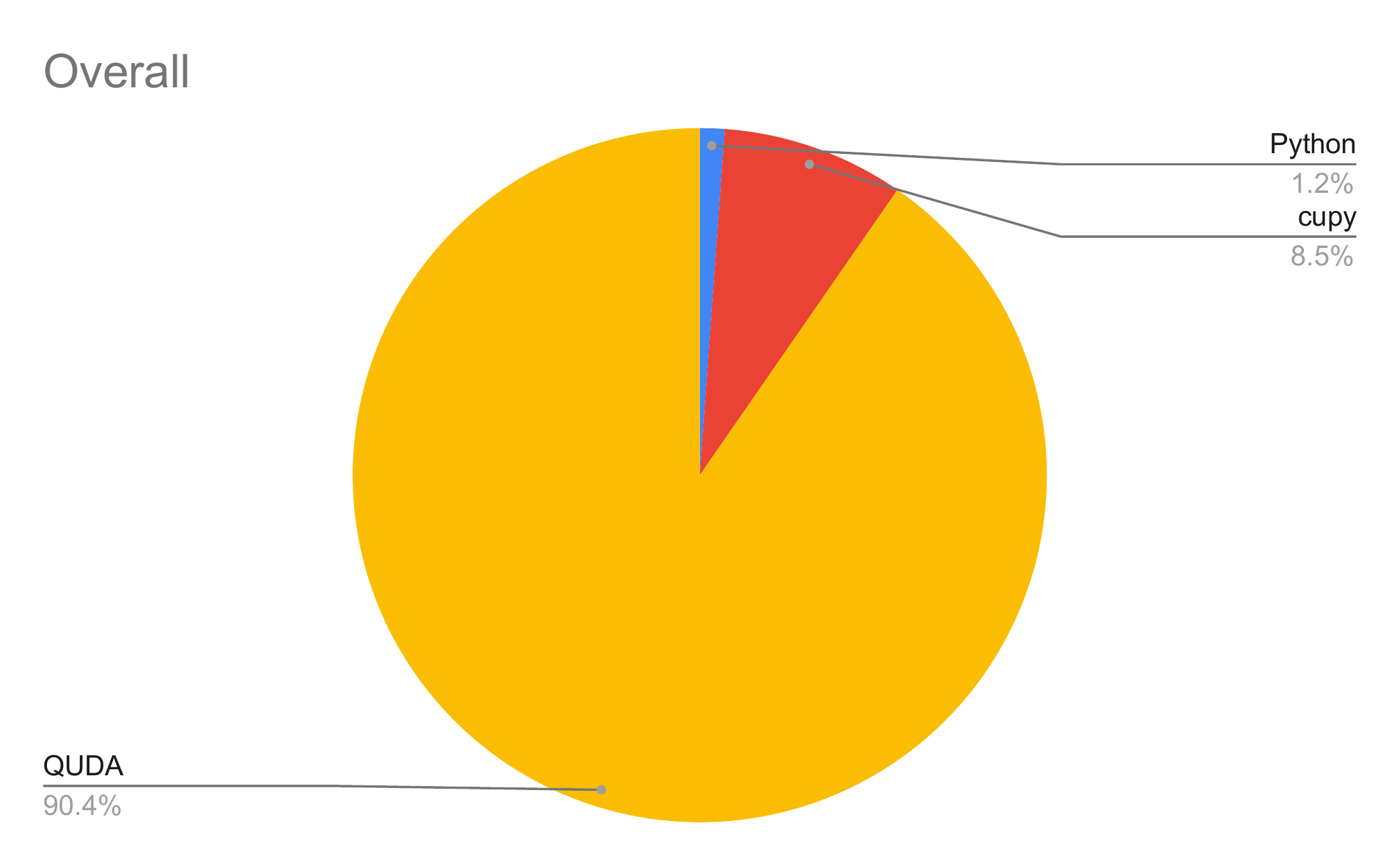}
\caption{The pie chart showing the relative proportion of the total computer time from each sector.}
\label{Fig:TotalPie}
\end{figure}

The breakdown of the overall performance is shown in Fig.~\ref{Fig:TotalPie}.  As can be seen in the figure, the largest contributor to the overall cost is the QUDA kernel.  This suggests most of the computation of the application program is handled by QUDA functions.  
The second largest contribution is from operations involving \texttt{CuPy} arrays, indicated by the label "cupy" in the pie chart.  Lyncs-QUDA uses \texttt{CuPy} arrays to represent lattice fields on GPU's such as gauge fields and the conjugate momentum fields internally, and some operations on these fields such as the calculation of momentum norm are handled on the Python side, not through QUDA.  As these operations are not directly related to the use of the Python interface but are part of the calculation for HMC simulation performed via \texttt{CuPy}, we separated its computational cost from the overhead due to usage of the Python interface and treat it in a similar manner to the computer time from QUDA kernels.
The cost due to the use of Python interface, on the other hand, is about 1.2\% of the total, which is marginal.

\begin{figure}
\centering
\begin{subfigure}{.5\textwidth}
  \centering
  \includegraphics[width=.75\linewidth]{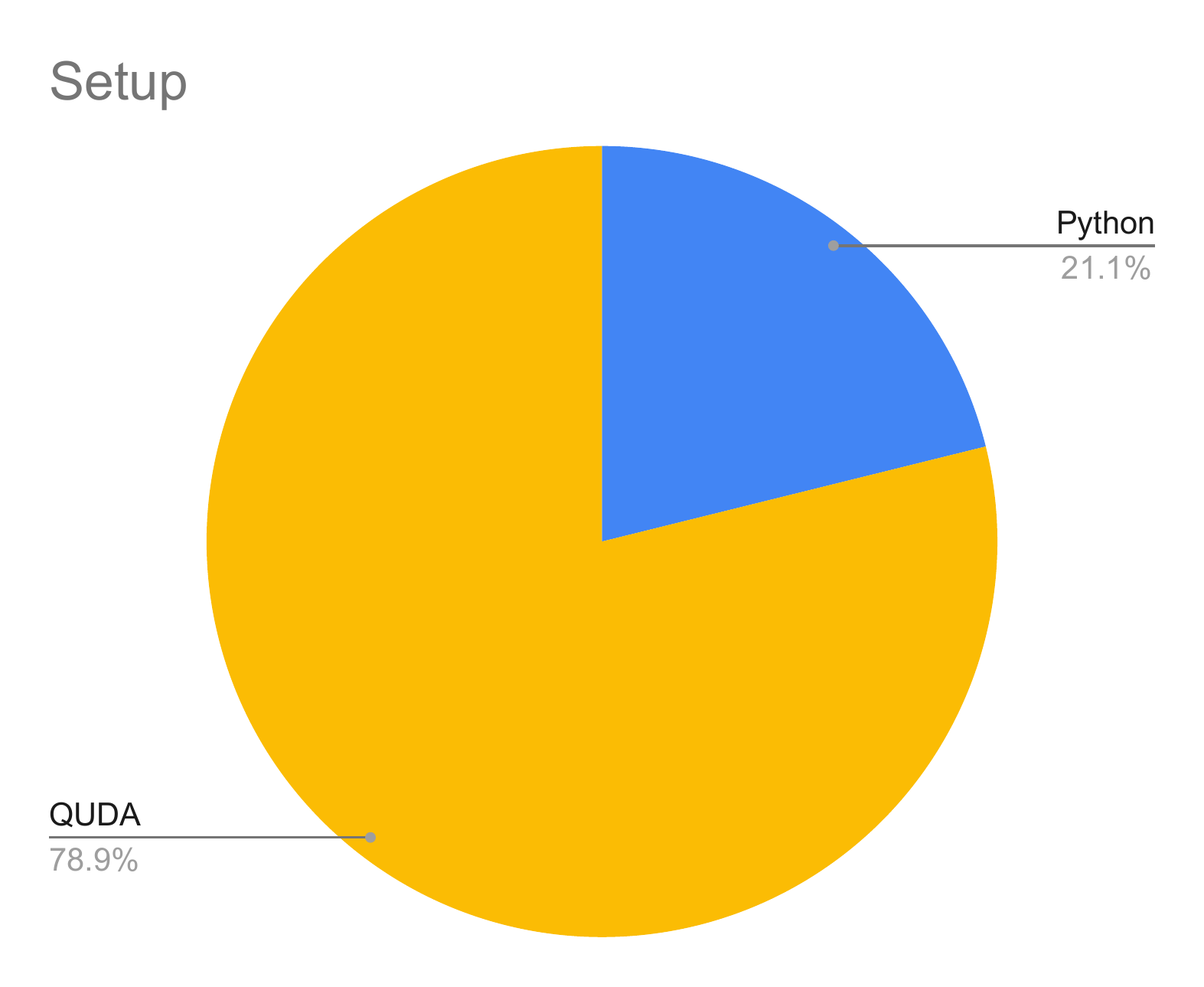}
  \caption{}
  \label{Fig:pieSetup}
\end{subfigure}%
\begin{subfigure}{.5\textwidth}
  \centering
  \includegraphics[width=\linewidth]{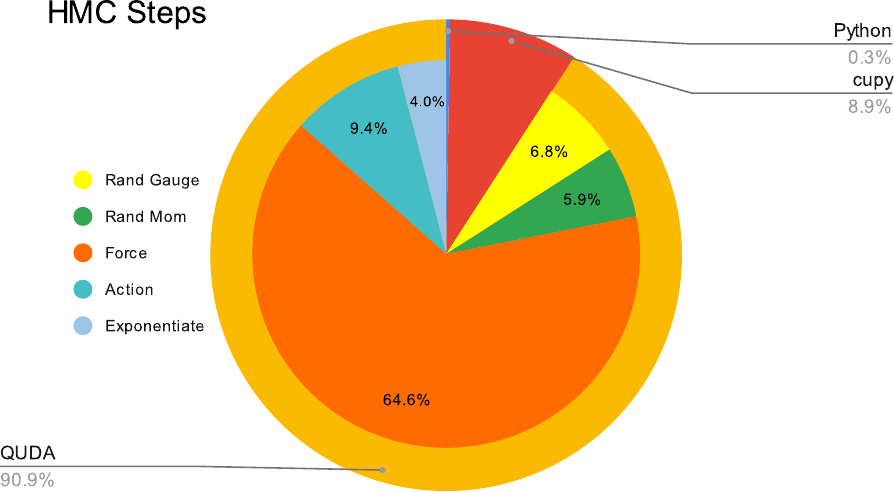}
  \caption{}
  \label{Fig:pieHMC}
\end{subfigure}
\caption{The pie charts showing the relative proportion of the computer time divided into the setup phase and HMC simulation phase.  The left plot is the pie chart showing the relative proportion of the computer time of the setup phase from each sector, and the right plot the pie chart showing the relative proportion of the computer time during HMC steps from each sector.}
\label{Fig:pieDivided}
\end{figure}

If we look more closely into the profiling result by breaking down the overall cost into the one from the setup phase and the other from repeating HMC steps, we find that the computer time for setup phase is 17.6s, which is about 4.1\% of the total time of 421s.  The pie chart for the setup phase is found in Fig.~\ref{Fig:pieSetup}.
Here again, the QUDA kernel occupies a major portion of the computer time, which is about 78.9\%.  Python accounts for 21.1\%.  The rest of the total computer time, which is about 95.8\% of the total, is coming from repetition of the HMC steps.  The pie chart for this phase can be found in Fig.~\ref{Fig:pieHMC}.   In this phase, we find that the overhead due to the usage of the Python interface is only about 0.3\%, and it is very small for this particular run.  Thus, most of the overhead of using the Python interface comes from the setup phase.  As setup is done only once at the beginning of the simulation, and as HMC steps are repeated in a large number of times in a practical simulation, not just one hundred times for this test run.  So the total overhead due to Python will be much smaller than shown in this example and will approach to 0.3\% in production runs.  

In Fig.~\ref{Fig:pieHMC}, the inner pie chart shows the breakdown of the computer time from the QUDA kernel.  As can be seen, the most computationally expensive part of the HMC steps is the computation of the gauge force for updating the gauge momentum.  Other components of the HMC simulation is not as computationally demanding as the force calculation.  Among those components, calculation of the gauge action dominated the computer time.  Generation of the initial random gauge and random conjugate momentum at each step is similar in terms of the computational cost, and exponentiation of the gauge momentum to obtain a new gauge configuration made up of the smallest part in the QUDA kernel.   


\subsection{Scaling}
\begin{figure}
\centering
\begin{subfigure}{.5\textwidth}
  \centering
  \includegraphics[width=\linewidth]{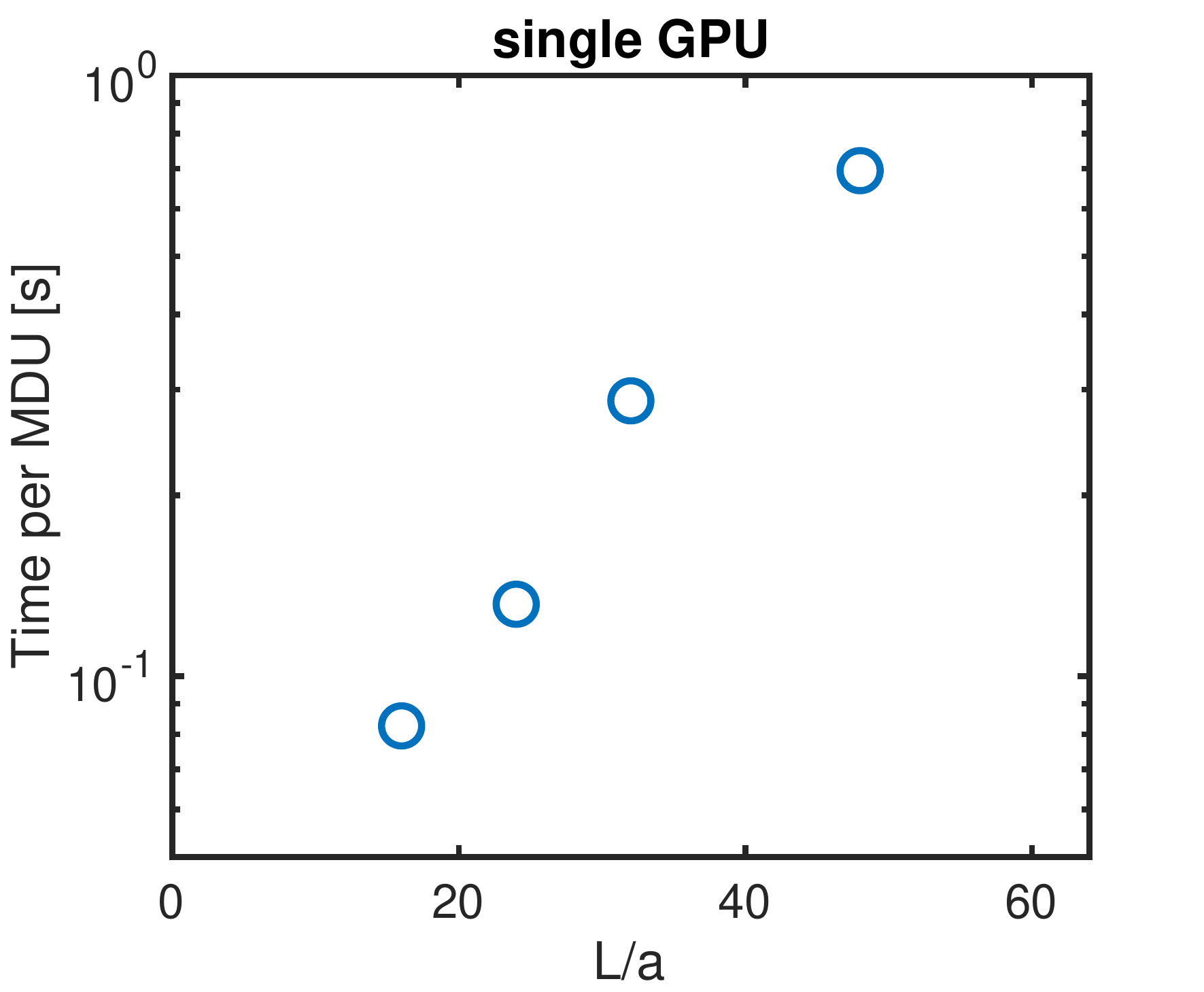}
  \caption{}
  \label{Fig:volumeScaling}
\end{subfigure}%
\begin{subfigure}{.5\textwidth}
  \centering
  \includegraphics[width=\linewidth]{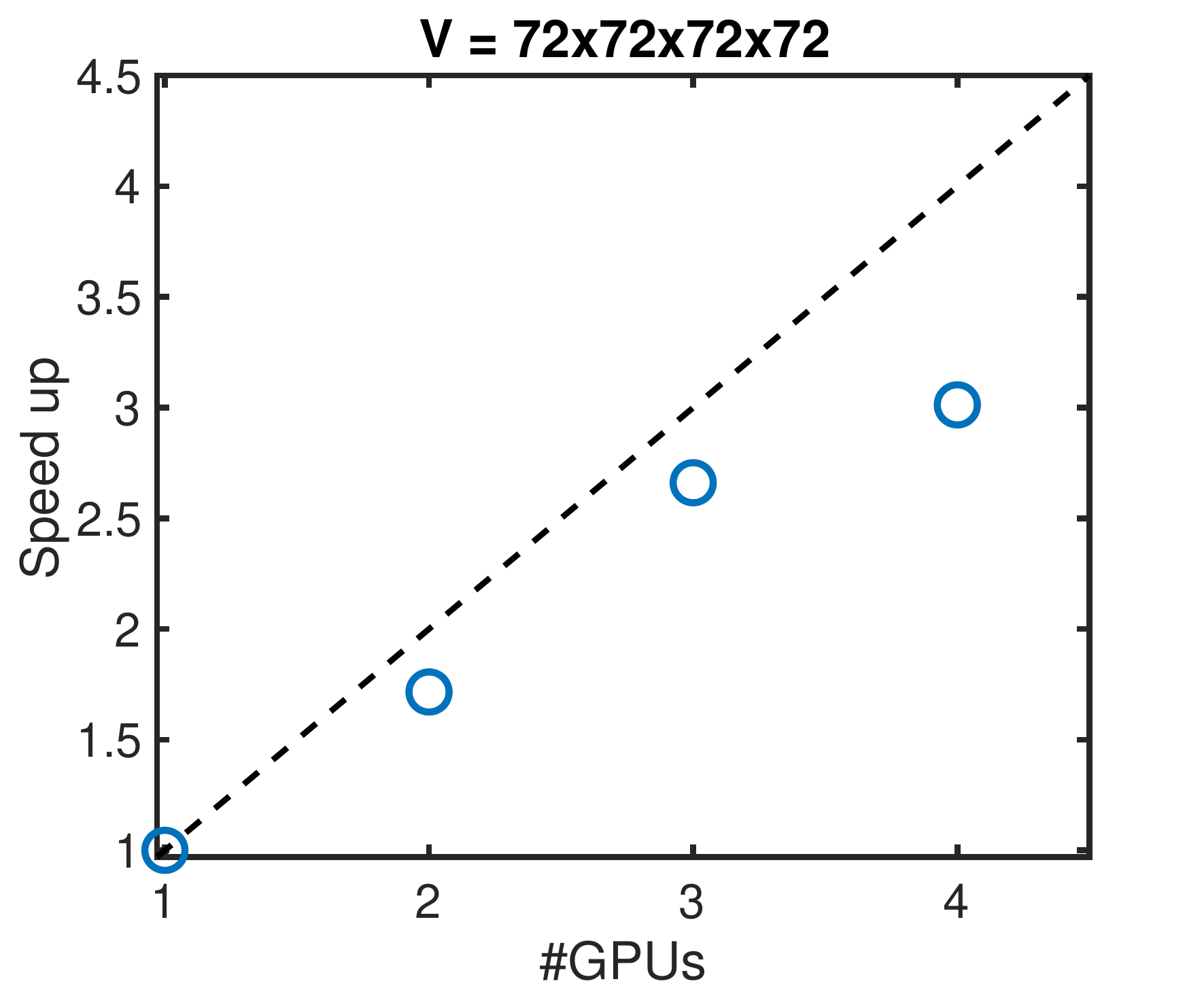}
  \caption{}
  \label{Fig:relSpeedup}
\end{subfigure}
\caption{The plot of the time per MDU in seconds as a function of the lattice spacing in lattice units, $a$ is shown on the left, and the plot showing speedup of the time per MDU with a different number of GPU's used relative to that of a single GPU on the right.}
\label{Fig:scaling}
\end{figure}

We have also preformed different scaling test.  This was done on JUWELS Booster at Jülich Supercomputing Centre, which comes with AMD EPYC 7402 and four GPU's (NVIDIA A100 Tensor Core) \cite{Krause:2019pey}.  Figure~\ref{Fig:scaling} shows two benchmarks.  Time per Molecular Dynamics Unit (MDU) as a function of the lattice spacing in the lattice unit, $\frac{L}{a}$, is shown in Fig.~\ref{Fig:volumeScaling}.  As can be observed, the time per MDU increases as a function of the lattice size.  Figure \ref{Fig:relSpeedup} shows the speedup of time per MDU with different number of GPU's used relative to the time with a single GPU, using a lattice of a fixed volume.  The speedup falls off from the ideal scaling as the number of GPU's used increases.

\section{Outlook}
Lyncs-API is designed to achieve performance, portability, and productivity by providing Python interfaces to many already existing libraries for lattice QCD calculations in various languages.  It is based on a number of other Python packages and implements extension to some of these to support implementation of the interfaces.  Lyncs-QUDA is one of such interfaces, in particular to QUDA.  Internally, it loads the QUDA library and uses it inside of various Python wrapper classes.  These classes exposes a number of QUDA objects and functions to provide the user a wider access to QUDA capabilities.

The example script performing HMC simulation showed that Lyncs-QUDA simplifies programming considerably and lets the user focus on the problem instead of the implementation details, while incurring insignificant overhead.  
Also, being a Python interface, it is straightforward to utilize other Python modules or various external Python packages. 
For example, we have used \texttt{cProfile} and \texttt{SnakeViz} in our profiling and visualization as well as \texttt{aim} for analyzing the simulation runs.  Scaling tests have also been conducted to see if the performance of our HMC script is reasonable, and the results confirm our expectation.  

This is a work in progress so that there are many features of QUDA that has not been incorporated into our interface.  They include but not limited to supporting different fermion types such as staggered fermion, supporting QUDA's various compiler options, and updating the version of QUDA on which the Python interface is based.  

\section*{Acknowledgement}
This project has received funding under PRACE-6IP, Grant agreement ID: 823767, Project name: LyNcs. LyNcs is one of 10 applications supported by PRACE-6IP, WP8 ``Forward Looking Software Solutions''. S.Y., S.B.~and J.F.~have received funding under this project.   The authors gratefully acknowledge the Gauss Centre for Supercomputing
e.V. (www.gauss-centre.eu) for funding computing time projects on the JSC supercomputer JUWELS Booster \cite{Krause:2019pey}.  We also thank the developers of the QUDA \cite{Clark:2009wm} library for their continued support.

\bibliographystyle{JHEP}
\bibliography{Lattice2022_SYamamoto}


\end{document}